# Harnessing Diet and Gene Expression Insights through a Centralized Nutrigenomics Database to Improve Public Health

Fahmida Hai[1], Shriya Samudrala[2], Ijeoma Ezengwa[2], Rubayat Khan[3], Saif Nirzhor[4], and Don Roosan[5]

[1]*Tekurai Inc., San Antonio, USA*
[2]*College of Health Sciences, Western University of Health Sciences, Pomona, USA*
[3]*University of Nebraska Medical Center, Omaha, USA*
[4]*University of Texas Southwestern Medical Center, Dallas, USA*
[5]*School of Engineering and Computational Sciences, Merrimack College, North Andover, USA*
fahmida@tekurai.com, samudralashriya@gmail.com, Ijeoma.ezengwa@gmail.com, rubayat.khan@unmc.edu, saif.nirzhor@utsouthwestern.edu, roosand@merrimack.edu

Keywords: Nutrigenomics, Nutrition, Public Health, Artificial Intelligence, Database Management

Abstract: Nutrigenomics is an emerging field that explores the intricate interaction between genes and diet. This study aimed to develop a comprehensive database to help clinicians and patients understand the connections between genetic disorders, associated genes, and tailored nutritional recommendations. The database was built through an extensive review of primary journal articles and includes detailed information on gene characteristics, such as gene expression, location, descriptions, and their interactions with diseases and nutrition. The data suggest that a patient's food intake can either increase or decrease the expression of genes related to specific diseases. These findings underscore the potential of nutrition to modify gene expression and reduce the risk of chronic diseases. The study highlights the transformative role nutrigenomics could play in medicine by enabling clinicians to offer personalized dietary recommendations based on a patient's genetic profile. Future research should focus on validating the database in clinical counselling to further refine its practical applications.

## 1 INTRODUCTION

The impact of nutrigenomics on daily life is profound and far-reaching. Nutrigenomics, the study of the relationship between diet and gene expression, is a rapidly evolving field with tremendous potential for growth, particularly through the use of advanced databases. Given the current health challenges faced by many communities, this area of research is increasingly relevant, as nutrition plays a pivotal role in the development and management of many chronic disorders, such as colorectal cancer. Nutrigenomics integrates diverse scientific and environmental factors to explore the intricate connections between nutrition and gene expression (Franzago et al, 2020). The origins of nutrigenomics trace back to the groundbreaking Human Genome Project of the late 20th century, which provided researchers with the tools to investigate how diet influences gene activity and contributes to the onset of diseases. As the field continues to grow, nutrigenomics holds the promise of advancing personalized nutrition and improving public health outcomes.

Nutrigenomics offers valuable insights into how individuals may respond differently to specific foods or nutrients based on their genetic makeup. Food serves as the primary source of nutrients essential for the body to function effectively, supporting daily activities and overall survival. Through nutrigenomics, individuals can better understand how modifying their diet can influence gene expression, particularly in cases involving cancerous tissues.

The database empowers consumers by providing knowledge to make informed dietary decisions that can positively impact their health. Research has identified numerous genes affected by dietary intake. For instance, in colorectal cancer, consuming more than 2 ounces of red meat per day has been linked to the downregulation of COL1A1, a gene involved in extracellular matrix (ECM) regulation and cell matrix adhesion (Zeng et al, 2023; Aykan, 2015). In addition to COL1A1, other genes such as the anti-metastasis and angiogenesis-related gene COL4A2, the tumor suppressor gene TP53, and the cytokine signaling regulation gene IL22RA1 have been implicated in cancer pathogenesis (Hall et al., 2016; Abo El-Ella & Bishayee, 2019; Nasir et al, 2019). Developing a deeper understanding of these interactions opens new opportunities for personalized dietary strategies to prevent and manage diseases, underscoring the transformative role of nutrition in promoting health and well-being.

Nutrigenomics also explores how nutrition interacts with our genes to influence brain function and mood, offering significant potential in the field of mental health (Marcum, 2020). As an emerging discipline, further research in nutrigenomics could lead to innovative treatments for mental health conditions, potentially minimizing the unpleasant side effects and withdrawal symptoms often associated with traditional medications. Key genes implicated in depression and anxiety include the serotonin transporter gene (SLC6A4/5HTT), the serotonin transporter-linked promoter region (5-HTTLPR), the serotonin receptor gene (HTR2A), and brain-derived neurotrophic factor (BDNF). These genes play critical roles in regulating mood and emotional well-being (Birla et al., 2022). Notably, L-tryptophan (TRP) supplements have shown promise in improving affective states by promoting serotonin synthesis, a neurotransmitter essential for mood regulation, appetite control, and sleep. Serotonin is synthesized from TRP, which can be obtained from foods such as salmon, nuts and seeds, turkey and poultry, and pineapple (Rodrigues et al., 2021). Identifying the genetic pathways influenced by nutrition, nutrigenomics leads to the potential to revolutionize mental health care, paving the way for personalized dietary strategies that enhance emotional and cognitive well-being.

Consumers have access to various companies that offer services to use genetic information to tailor dietary recommendations to their unique genome. However, the databases currently available tend to be more generalized, offering insights into genes associated with common diseases rather than personalized, specific guidance (Jaskulski et al, 2023). Among these, NutrigenomeDB stands out as a primary resource for finding nutrition-specific articles related to various diseases. While other databases exist, their focus is often limited to the general connection between diet and genes in the field of nutrigenomics. To improve the functionality and precision of these databases, several key concepts should be incorporated. First, databases should specify the required intake amounts of specific foods necessary to influence gene expression effectively. Additionally, providing detailed descriptions of the interactions between particular foods and genes, along with the precise location of these genes, would enhance their utility for consumers and researchers alike (Jaskulski et al, 2023).

As a relatively new and emerging field, nutrigenomics faces challenges in accessing comprehensive information on the effects of nutrition on gene expression. While existing studies have explored a variety of nutrigenomic interactions, there remains a lack of literature that delves deeply into these interactions on a disease-specific level (Alegría-Torres et al, 2011). Continued research is essential to bridge this gap, as nutrigenomics has the potential to significantly advance our understanding of these mechanisms and contribute to the development of targeted interventions and treatments. With increased funding and resources, we anticipate a surge in scientific studies and research in the near future, enabling nutrigenomics to fulfill its potential as a transformative tool in personalized medicine and nutrition.

There is a pressing need for companies to provide more detailed insights into how specific interactions between food and genes result in positive health outcomes. It is essential for consumers to have the ability to track and understand the pathways involved in these interactions, enhancing their knowledge of how nutrigenomics works. Given the strong correlation between nutrigenomics and diseases, understanding the effects of excessive or insufficient intake of certain foods is critical. For instance, increased consumption of certain foods can lead to epigenetic changes, such as methylation, which can alter gene sequences. An example is the overconsumption of red meat, which has been linked to the downregulation of the NCL gene in colorectal cancer (Genkinger & Koushik, 2007). Conversely, improper portion sizes or poor dietary habits can lead to the overexpression or underexpression of genes, further contributing to disease risk (Zhang et al, 2019).

With the rising incidence of hospital admissions related to poor nutrition and the rapid advancements in nutrigenomics, there is hope that personalized dietary strategies can help combat chronic diseases linked to nutrition (Peña-Romero et al, 2017). By emphasizing proper food intake and understanding the genetic implications of diet, nutrigenomics has the potential to transform healthcare and reduce the burden of nutrition-related illnesses. Recognizing the profound impact of nutrients on gene expression and overall health has the potential to enhance the well-being of entire communities, driving a new era of precision health and disease prevention.

## 2 METHODS

We identified a suitable database, Nutrigenetics, to collect data on gene classification and the interaction between genes and specific nutrients. The Nutrigenetics database provides information on various gene-food interactions and allowed us to access relevant articles related to specific nutrients (Martín-Hernández et al, 2019). During the selection process, shown in Figure 1, we noted the database's limitations, including a lack of detailed explanations on gene-nutrition interactions and recommended intake amounts. These factors were considered as part of our inclusion criteria. A total of 196 articles were thoroughly reviewed to extract data on the effects of nutrients on genes associated with chronic diseases. We carefully examined the limitations and concerns highlighted by researchers in the "gap" sections of these articles, identifying commonalities among them. To ensure the reliability of the findings, we analyzed the methods and results sections of each study to confirm the presence of robust and supporting data. Based on the criteria listed in Table 1, the pool of articles was narrowed down to eight. Two independent reviewers conducted the data extraction process, entered the extracted information into the database, and cross-checked all inputs for accuracy. This rigorous approach ensured the integrity and reliability of the data collected for our study.

Gene location was included in the database to provide patients with detailed information about where each gene is positioned on their chromosomes. This addition enhances patients' understanding of their genetic makeup. Gene descriptions were also incorporated to explain each gene's function in its natural state, uninfluenced by specific nutrients. This information helps patients grasp the potential mechanisms through which certain foods may contribute to cancer development.

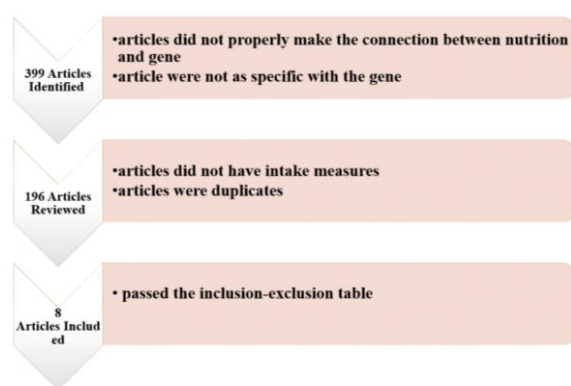

Figure 1: Flow chart illustrating the process of paper selection for inclusion in the study.

Table 1: The inclusion and exclusion criteria were implemented to create an adequate and proper nutrigenomics database.

| Inclusion Criteria | Exclusion Criteria |
|---|---|
| Articles available in full text | Articles is not available in full text |
| Articles has correlation between the nutrition and the gene | Articles does not have correlation between the nutrition and the gene |
| Articles have measure of consumption, whether that be numerical intake or frequency intake. | Articles does not have measure of consumption, whether that be numerical intake or frequency intake. |
| Method and result section of the paper has proper and supporting data | Method and result section of the paper does not have proper and supporting data |

Food intake was measured using two different approaches. The first method quantified the nutrient intake in numeric values. The second method relied on a Food Frequency Questionnaire (FFQ) to assess consumption frequency. This questionnaire included 11 response categories, ranging from "never" to "2 or more times a day," and reflected dietary habits from the year prior to the study.

Data collected from various research articles and studies were initially compiled into an Excel sheet and subsequently transformed into a comprehensive database. The database was implemented using MySQL, a widely used relational database management system, with data structured into tables for genes, diseases, nutrients, and their interactions, allowing for efficient querying and retrieval of information.

# 3 RESULTS

The database results revealed critical insights into the correlation between various nutritious foods and gene expression. It offers a comprehensive depiction of gene characteristics and their interactions with

By integrating these elements, the database equips clinicians and patients with a comprehensive understanding of the intricate relationships between nutritious foods, gene expression, and disease prevention. This resource has the potential to inform personalized nutrition strategies and enhance efforts to prevent and manage chronic diseases. To evaluate the functionality and utility of our database, we

| Diseases | P-value | Gene | Gene Expression | Location | Species | Description | Function | Interaction |
|---|---|---|---|---|---|---|---|---|
| Colorectal Cancer | 1.09E-04 | COL1A1 | https://www.ncbi.nlm.nih.gov/gene/1277 | 7q21.33 | Homo Sapiens | MHC class II subcluster, a 'stromal cluster | involved in the regulation of the ECM and cell matrix adhesion | % high intake of red meat correlated with downregulation |
| Colorectal Cancer | 1.22E-04 | COL1A2 | https://www.ncbi.nlm.nih.gov/gene/1278 | 7q21.3 | Homo Sapiens | MHC class II subcluster, a 'stromal cluster | involved in the regulation of the ECM and cell matrix adhesion | high intake of red meat correlated with downregulation |
| Colorectal Cancer | 6.44E-05 | Col4a2 | https://www.ncbi.nlm.nih.gov/gene/1284 | 13q34 | Homo Sapiens | Collagen-related genes; Upregulation anti- metastasis and angiogenesis | | high intake of red meat correlated with downregulation |
| Colorectal Cancer | 1.39E-06 | TIMP1 | https://www.ncbi.nlm.nih.gov/gene/7076 | Xp11.3 | Homo Sapiens | MHC class II subcluster, a 'stromal cluster; tissue inhibitor | involved in the regulation of the ECM and cell matrix adhesion | high intake of red meat correlated with downregulation, Diet downregulation |
| Colorectal Cancer | 8.19E-06 | SPARC | https://www.ncbi.nlm.nih.gov/gene/6678 | 5q33.1 | Homo Sapiens | MHC class II subcluster, a 'stromal cluster | important role in bone health, with meat protein increasing acid load and regulating calcium balance | high intake of red meat correlated with downregulation of calcium-binding genes and meat consumption modulates bone metabolism |
| Colorectal Cancer | 6.32E-05 | FZD5 | https://www.ncbi.nlm.nih.gov/gene/7976 | 8p21.1 | Homo Sapiens | which consists of genes that encode seven-transmembrane domain proteins. These proteins act as receptors for the signaling proteins of the wingless type MMTV integration site family. | assembly and structure, calcium binding, and the regulation of signal transduction through the Wnt signaling pathway. | high intake of red meat correlated with downregulation |
| Colorectal Cancer | 1.56E-04 | TMEM132B | https://www.ncbi.nlm.nih.gov/gene/114795 | 12q24.31-q24.32 | Homo Sapiens | It is anticipated that this will be an essential part of the membrane. | 1. assembly and structure 2. calcium binding 3. regulation of signal transduction through the Wnt signaling pathway. | high intake of red meat correlated with downregulation |
| Colorectal Cancer | 3.22E-04 | CCDC150 | https://www.ncbi.nlm.nih.gov/gene/284992 | 2q33.1 | Homo Sapiens | Protein coding | 1. assembly and structure 2. calcium binding 3. regulation of signal transduction through the Wnt signaling pathway. | high intake of red meat correlated with downregulation |
| Colorectal Cancer | 3.86E-05 | NED1 | https://www.ncbi.nlm.nih.gov/gene/85407 | 16q12.1 | Homo Sapiens | protein coding | | high intake of red meat correlated with downregulation |
| Colorectal Cancer | 4.87E-05 | TP53 | https://www.ncbi.nlm.nih.gov/gene/7157 | 17p13.1 | Homo Sapiens | Tumor suppressor gene | represses encoding of P53 | high intake of red meat correlated with downregulation |
| Colorectal Cancer | 3.03E-04 | THY1 | https://www.ncbi.nlm.nih.gov/gene/7070 | 11q23.3 | Homo Sapiens | Regulation anti- metastasis and angiogenesis | associated acute oxidative stress and delayed cytotoxic stress | high intake of red meat correlated with downregulation |
| Colorectal Cancer | 1.03E-04 | NCL | https://www.ncbi.nlm.nih.gov/gene/4691 | 2q37.1 | Homo Sapiens | Regulation anti- metastasis and angiogenesis | associated acute oxidative stress and delayed cytotoxic stress | high intake of red meat correlated with downregulation |
| Colorectal Cancer | 7.40E-06 | KRT1 | https://www.ncbi.nlm.nih.gov/gene/3848 | 12q13.13 | Homo Sapiens | Regulation anti- metastasis and angiogenesis | associated acute oxidative stress and delayed cytotoxic stress | high intake of red meat correlated with downregulation |
| Colorectal Cancer | 6.44E-05 | SLC17A4 | https://www.ncbi.nlm.nih.gov/gene/10050 | 6p22.2 | Homo Sapiens | Phosphate transport regulation gene | Regulation of phosphate transport | high intake of red meat correlated with upregulation |
| Colorectal Cancer | | SLC25A25 | https://www.ncbi.nlm.nih.gov/gene/114789 | 9q34.11 | Homo Sapiens | Phosphate transport regulation gene | Regulation of phosphate transport | high intake of red meat correlated with upregulation |
| Colorectal Cancer | 2.12E-04 | MUC17 | https://www.ncbi.nlm.nih.gov/gene/140453 | 7q22.1 | Homo Sapiens | The gene that regulates cytokine signaling is associated with the expression of specific types of cancer. | Regulation of cytokine signaling; downregulation play a role in meat-induced inflammation and inflammation-related diseases. | high intake of red meat correlated with upregulation |
| Colorectal Cancer | 1.17E-04 | IL22RA1 | https://www.ncbi.nlm.nih.gov/gene/58985 | 1p36.11 | Homo Sapiens | The cytokine signaling regulation gene is synthesized by T | The downregulation of cytokine signaling is believed to contribute to meat-induced | high intake of red meat correlated with downregulation of calcium-binding genes and meat consumption modulates bone metabolism |
| Colorectal Cancer | 3.66E-05 | CEMIP | https://www.ncbi.nlm.nih.gov/gene/57214 | 15q25.1 | Homo Sapiens | calcium-binding gene | Meat protein has a significant impact on bone health by increasing acid load and regulating calcium balance. | high intake of red meat correlated with downregulation of calcium-binding genes and meat consumption modulates bone metabolism |
| Colorectal Cancer | 8.23E-05 | HSP45 | https://www.ncbi.nlm.nih.gov/gene/3309 | 9q33.3 | Homo Sapiens | calcium-binding gene | Meat protein has a significant impact on bone health by increasing acid load and regulating calcium balance. | high intake of red meat correlated with downregulation of calcium-binding genes and meat consumption modulates bone metabolism |
| Colorectal Cancer | 3.50E-04 | PVDE | https://www.ncbi.nlm.nih.gov/gene/221808 | 7p21.3 | Homo Sapiens | calcium-binding gene | Meat protein has a significant impact on bone health by increasing acid load and regulating calcium balance. | high intake of red meat correlated with downregulation of calcium-binding genes and meat consumption modulates bone metabolism |
| Colorectal Cancer | 2.09E-04 | C1R | https://www.ncbi.nlm.nih.gov/gene/715 | 12p13.31 | Homo Sapiens | calcium-binding gene | Meat protein has a significant impact on bone health by increasing acid load and regulating calcium balance. | high intake of red meat correlated with downregulation of calcium-binding genes and meat consumption modulates bone metabolism |
| Colorectal Cancer | 3.78E-06 | CLDN1 | https://www.ncbi.nlm.nih.gov/gene/9076 | 3q28 | Homo Sapiens | calcium-binding gene | Meat protein has a significant impact on bone health by increasing acid load and regulating calcium balance. | high intake of red meat correlated with downregulation of calcium-binding genes and meat consumption modulates bone metabolism |
| Colorectal Cancer | 1.30E-04 | EGFLAM | https://www.ncbi.nlm.nih.gov/gene/133584 | 5p13.2-p13.1 | Homo Sapiens | calcium-binding gene | Meat protein has a significant impact on bone health by increasing acid load and regulating calcium balance. | high intake of red meat correlated with downregulation of calcium-binding genes and meat consumption modulates bone metabolism |
| Colorectal Cancer | 2.83E-04 | TRIL | https://www.ncbi.nlm.nih.gov/gene/9865 | 7p14.3 | Homo Sapiens | genes associated with immune response | | Downregulated by high meat consumption |

Figure 2: Snapshot of a section of the database, showcasing detailed information on the relationship between nutritious foods, gene expression, and disease prevention. This resource offers valuable insights for both clinicians and patients.

specific foods. The data demonstrated that the intake of certain foods can either upregulate or downregulate the expression of genes associated with specific diseases. The database is organized into ten sections: disease, associated gene, P-value, gene expression, location, species, description, function, interaction, and intake amount. Each gene is linked to its corresponding disease, along with the P-value that quantifies the strength of its interaction with particular foods. The inclusion of P-values provides a clear understanding of the statistical significance of these correlations.

The gene expression section highlights the gene's characteristics, including its type and a summary of how it is expressed. The location section specifies the gene's position on the chromosome, with the first number representing the chromosome, the "p" indicating the short arm, and the "q" denoting the long arm. The database also includes concise descriptions of each gene and their biological functions.

The interaction section outlines how specific foods interact with genes to influence their expression, offering insights into the metabolic pathways involved. The intake amount category provides a quantitative measure of the required nutrient consumption to induce changes in gene expression.

conducted several test queries; for example, querying the database for genes associated with colorectal cancer and their interaction with red meat consumption returned results including the downregulation of COL1A1 with a recommended intake limit of less than 2 ounces per day. In contrast, a similar query in NutrigenomeDB provided general information on gene-diet interactions but lacked specific intake recommendations, highlighting the enhanced detail our database offers. To demonstrate the database's implementation and functionality, we present a sample query output in Figure 3, which shows the detailed information retrieved for the gene MTHFR and its interaction with folate, including the gene's location, description, and the specific intake amount that influences its expression.

# 4 DISCUSSIONS

Nutrigenomics is an emerging and rapidly evolving field, but current research remains limited and fragmented. While our database underscores the significant impact of nutrition on gene expression and its potential to influence disease risk, it is crucial to acknowledge that nutrition is one of several factors contributing to chronic disease management. Genetic predispositions, medical treatments, and lifestyle choices all play integral roles, and thus, the nutritional recommendations provided should be considered as part of a holistic healthcare approach. While valuable information exists, compiling a cohesive and comprehensive understanding of disorders, their associated genes, and corresponding nutritional recommendations often requires navigating multiple, often inaccessible, sources. To address this gap, we have developed a curated database that systematically organizes diseases, associated genes, and specific nutritional interventions that can upregulate or downregulate these genes.

This database serves as a valuable resource for nutrition counselors and healthcare clinicians, offering evidence-based recommendations to support patient health. Currently, our research focuses on specific cancers and neuropsychiatric disorders—relatively new areas of study within nutrigenomics. As the field continues to grow, we anticipate an expanding body of knowledge that will enable the identification of additional targets and further enhance personalized approaches to nutrition and disease prevention.

The potential applications of this database extend across various domains, including nutritional counselling, informatics, and public health. Insights derived from the database enable clinicians to offer personalized, gene-based dietary recommendations tailored to an individual's unique genetic profile (Agrawal et al, 2023; Mullins et al, 2020). Such personalized nutrition plans not only enhance the effectiveness of dietary interventions but also reduce the risk of nutrition-related diseases, presenting a transformative approach to preventive healthcare (Meng et al., 2019).

Current research in nutrigenomics has significant implications for informatics. The database serves as a centralized, well-organized resource that researchers can easily access to support the discovery of new genes and dietary interventions. Moreover, the nutrigenomics database can integrate with advanced technologies, such as artificial intelligence (AI), to unlock deeper insights from large datasets of genetic and nutritional information. AI applications have the potential to analyze these datasets, identify patterns and associations, and develop predictive models to assist healthcare clinicians in making informed decisions about nutrition and health (Gao & Chen, 2017; Marcum, 2020). Currently, healthcare applications utilizing AI, augmented and virtual reality have been increasing rapidly. This technology is being utilized for patient counselling, medication reminders, and even to aide in surgery (Roosan, 2024; Li et al., 2023). For example, AI could generate personalized nutrition plans tailored to an individual's unique genetic profile that the provider reviews (Roosan, 2024). By analyzing genetic data and identifying variations that influence how an individual metabolizes specific nutrients or responds to certain diets, AI can provide highly targeted dietary recommendations (Malle, 2021). This integration of AI with nutrigenomics could revolutionize the field, paving the way for more precise and effective healthcare solutions.

Integrating the nutrigenomics database with electronic health records (EHRs) could significantly enhance its clinical utility. With access to comprehensive patient data, including genetic profiles and dietary habits, clinicians can develop more precise and tailored nutrition recommendations (Roosan et al, 2019).

The applications of genomic knowledge extend far beyond healthcare. Genomics forms the foundation for "personalized medicine," offering unique, patient-specific clinical interventions that improve outcomes. Beyond healthcare, genomics has transformative potential in agriculture, environmental science, forensic science, and drug development (Senthil et al., 2019).

From a public health perspective, the database could enhance the effectiveness of interventions by providing tailored nutritional recommendations based on genetic profiles and dietary needs (Alegría-Torres et al, 2011; Roosan et al, 2024). Targeted strategies could address specific subpopulations with unique genetic traits and nutritional risk factors, enabling the development of more effective public health policies and programs (Wu et al., 2024; Roosan et al, 2024). The database also provides valuable insights into the relationships between genetics, nutrients, and health outcomes, offering opportunities to identify therapeutic targets and improve disease prevention at the population level (Fenech et al., 2011).

There is growing evidence linking genetics, poor nutrition, and disease risk. For example, individuals with genetic variations affecting folate metabolism are at an increased risk of colorectal cancer when consuming low-folate diets (Kim, 2006; Zeng et al, 2023). Variations in the MTHFR gene can impair folic acid metabolism, and excessive folic acid consumption in these individuals may mask vitamin B12 deficiency, leading to adverse health outcomes (Birla et al., 2022). Diets high in processed and red meat have similarly been associated with a heightened risk of colorectal cancer, particularly in individuals with genetic predispositions (Aykan, 2015; Bertucci et al, 2004). Additionally, genetic variations have been shown to influence susceptibility to mental health disorders such as depression and anxiety, with poor dietary quality exacerbating these conditions. These findings highlight the intricate interplay between genetics, nutrition, and health, emphasizing the importance of personalized dietary strategies to mitigate risks and improve outcomes.

The database is designed primarily for clinicians, offering a robust tool to assist in creating personalized treatment plans. By analyzing genetic profiles, clinicians can identify variations that influence nutrient metabolism and recommend tailored dietary changes or supplements to optimize patient health. Equally important is educating patients about the benefits of the database, which can be achieved through online resources and consultations with healthcare professionals. Empowering patients with this knowledge enables them to take an active role in managing their health through informed dietary and lifestyle changes (Roosan et al, 2022; Roosan, 2022).

It is important to note that while nutrigenomics provides valuable insights into potential areas of concern and personalized strategies, it is not a diagnostic tool. Instead, it serves as a resource to enhance understanding and promote proactive health management. Although nutrigenomics is still in its early stages, it represents the beginning of a deeper understanding of the complex relationship between genetics and nutrition. This field has already made significant strides in identifying how genetic factors and nutrients interact to impact health and reduce disease risk. As research continues to advance, we anticipate a growing body of knowledge that will pave the way for more personalized approaches to medicine, tailoring healthcare interventions to individual genetic profiles.

While many companies currently focus on genomic testing to raise awareness of genetic diseases, there is a growing need for these companies to provide customers with more comprehensive information on the connection between genes and nutrition. Enhancing access to such insights would enable customers to develop a deeper understanding of nutrigenomics. Achieving this goal may involve initiatives such as education programs, accessible genetic testing, consultations with healthcare providers, and greater transparency from companies (Guasch-Ferré et al., 2018). These studies collectively highlight the multifaceted nature of clinical decision-making, exploring cognitive strategies, complexity measurement, expert heuristics, and disease-specific reasoning to inform the design of effective decision support systems ((Islam et al, 2014; Islam et al, 2015; Islam et al, 2016a; Islam et al, 2016b). By expanding awareness and accessibility, nutrigenomics has the potential to transform how individuals approach health and nutrition, fostering a deeper understanding of personalized wellness.

## 4 CONCLUSIONS

In conclusion, the nutrigenomics database offers a comprehensive understanding of the relationship between nutritious foods, gene expression, and disease prevention. By cataloging diseases, associated genes, and specific nutritional interventions that can upregulate or downregulate gene activity, this resource serves as a valuable tool for nutrition counselors and healthcare clinicians seeking evidence-based recommendations to support their patients' health. The database has far-reaching implications for nutritional counseling, informatics, and public health. It holds the potential to enable the creation of personalized nutrition plans tailored to an individual's genetic makeup, integrate seamlessly with AI applications for data analysis and predictive modeling, and enhance clinical practice through integration with EHRs.

Furthermore, genomics knowledge provides opportunities for personalized medicine, agriculture environment, forensic science, and drug development. The nutrigenomics database plays a crucial role in preventing or managing disease through diet and recognizing the significant impact of nutrients on overall health, which can notably enhance the health and well-being of the entire community. Lastly, the nutrigenomics database could help identify targeted interventions for subgroups of the population with specific genes and risk factors of nutritional needs and offer valuable insights into disease pathogenesis and potential targets for interventions, ultimately optimizing health in individuals and communities through nutrigenomics.

## ACKNOWLEDGEMENTS

We are grateful to Merrimack College for support.